\documentclass[runningheads]{llncs}
\usepackage{graphicx}
\usepackage{svg}
\usepackage{amsmath}
\usepackage{booktabs} 
\usepackage{tabularx}
\usepackage{makecell} 
\usepackage{amsmath}  
\usepackage{algorithm}
\usepackage{algpseudocode}
\usepackage{subfigure}
\usepackage{todonotes} 

\begin{document}
%
\title{A Novel Technique for Query Plan Representation Based on Graph Neural Nets}
%
%


\author{Baoming Chang\inst{1}\orcidID{0009-0005-7414-3631}
\and
Amin Kamali\inst{1}\orcidID{0000-0003-2176-4088}
\and
Verena Kantere\inst{1}\orcidID{0000-0002-3586-9406}}
\authorrunning{B. Chang et al.}
\titlerunning{A Novel Technique for Query Plan Representation Based on GNNs}
%
\institute{University of Ottawa, 75 Laurier Ave E, Ottawa, ON K1N 6N5, Canada
\email{\{bchan081,skama043,vkantere\}@uottawa.ca}}
\maketitle


\begin{abstract}
Learning representations for query plans play a pivotal role in machine learning-based query optimizers of database management systems. To this end, particular model architectures are proposed in the literature to transform the tree-structured query plans into representations with formats learnable by downstream machine learning models. However, existing research rarely compares and analyzes the query plan representation capabilities of these tree models and their direct impact on the performance of the overall optimizer. To address this problem, we perform a comparative study to explore the effect of using different state-of-the-art tree models on the optimizer's cost estimation and plan selection performance in relatively complex workloads. Additionally, we explore the possibility of using graph neural networks (GNNs) in the query plan representation task. We propose a novel tree model BiGG employing \underline{Bi}directional \underline{G}NN aggregated by \underline{G}ated recurrent units (GRUs) and demonstrate experimentally that BiGG provides significant improvements to cost estimation tasks and relatively excellent plan selection performance compared to the state-of-the-art tree models.

\end{abstract}
\section{Introduction}

Query optimization is a critical component of a database management system due to its difficulty and importance in query execution performance. The process of accurately and efficiently estimating the cost of generated candidate query execution plans and selecting the optimal plan is always a challenge in the field of query optimization. A query execution plan is typically represented as a tree, where the nodes contain information about operators used to access, join, or aggregate data, and the edges contain dependencies between the parent and child nodes. An optimal plan tree enables the database management system to access and manipulate data efficiently. Traditional query optimizers employ cost models to estimate the amount of processing data using plan trees. These cost models rely heavily on statistical methods such as Histograms~\cite{histogram}, which are susceptible to large errors because of their inability to effectively capture characteristics such as join-crossing correlations.

\begin{figure}[t]
\centering
\includegraphics[width=\textwidth]{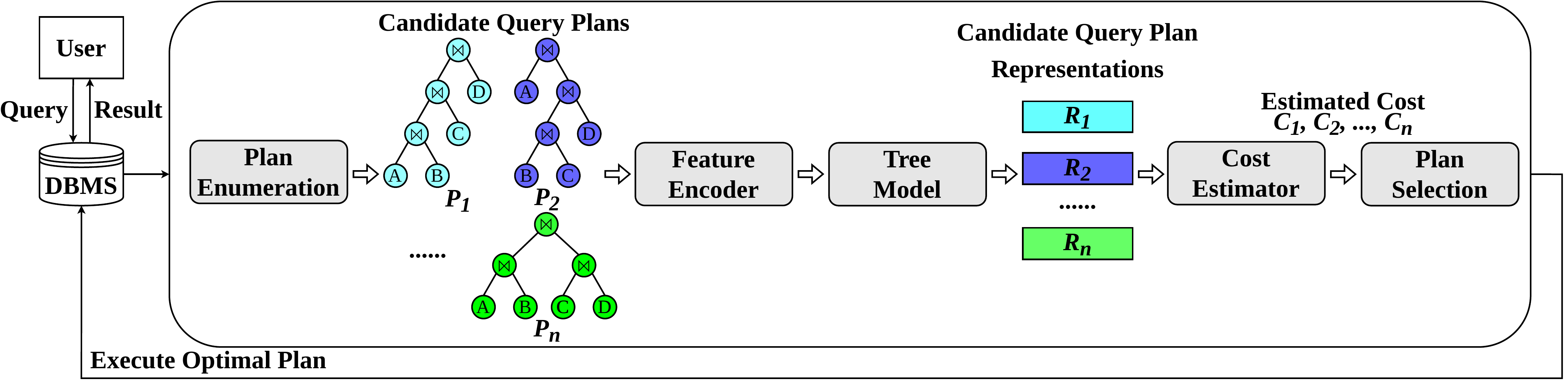}
\caption{Machine learning-based query optimizer framework.} \label{ML Query Optimizer Framework}
\vspace {-0.15in}
\end{figure}

The development of machine learning provides a promising solution to improve query optimization and has been proposed in a wide range of applications in this field. Recent machine learning-based query optimizers employ a similar structure, as shown in Fig.~\ref{ML Query Optimizer Framework}. In these frameworks, node information from a query plan tree is first encoded into node features. Then, a tree model transforms and aggregates these into graph-level representations for each candidate query plan, and a cost estimator predicts execution costs based on these representations, enabling the optimizer to select the most efficient plan. The challenge in this process is to effectively transform the tree-structured query plan into a graph-level vector while preserving as much of the original node features and structural information as possible. The quality of representation generated by the tree model is crucial to ensure the cost estimator model makes accurate predictions and allows the optimizer to make informed decisions for specific tasks involving plan selection, such as join order optimization and query rewriting.

Current research in query optimization often lacks direct performance comparisons of the representation abilities of these tree models, focusing instead on the general performance of the whole query optimization process~\cite{2023comparative}. Therefore, in this paper, we evaluate the mainstream tree models query plan representation under more complex workloads, comparing and analyzing their performance in cost estimation tasks. In addition, unlike previous studies~\cite{2023comparative}, instead of pairwise comparisons, we compare and analyze the tree model's plan selection capability in a way that is closer to actual usage scenarios by selecting an optimal plan from multiple candidate query plans to explore how different tree models affect the performance of plan selection.

Our research also explores the possibility of using graph neural networks (GNNs)~\cite{gnn} as tree models. Although GNNs are being developed and have achieved success across various graph-based domains, their application within query plan cost estimation remains insufficiently explored. In this context, our paper explores the potential of employing GNNs for query execution plan representation learning. We also propose a novel tree model architecture for query plan representation based on bidirectional GNN~\cite{dirgnn} and a GRU-based aggregation method~\cite{gruaggr}, which can capture the intricacies of query plan trees more accurately and robustly, setting a new stage for query plan representation learning.

To summarize, our main contributions are:
\begin{itemize}
    \item We conduct a comparative analysis of the performance of mainstream tree models in cost estimation under complex conditions.
    \item We evaluate the impact of tree models in the task of selecting the optimal plan from multiple candidate plans, closely mirroring actual application scenarios, and provide an intuitive comparison and analysis.
    \item We thoroughly explore the possibility of using GNNs as a query execution plan tree model.
    \item We propose a novel query plan tree model BiGG based on bidirectional GNN and GRU, demonstrating its superior representational capabilities and performance compared to existing tree models.
\end{itemize}

In the rest of the paper: Section 2 presents the problem of query plan representation and current tree models. Section 3 outlines related works along with their merits and shortcomings. Section 4 introduces the proposed architecture of BiGG. Section 5 presents the comparative study of existing and GNN-based tree models' performance on cost estimation and plan selection tasks, and Section 6 concludes the paper and discusses future work.

\section{Problem Statement}
In a query optimizer based on machine learning, query plan representation is a module that takes the physical query plan as input and uses a feature encoder and a tree model to generate vectors that subsequent machine learning models use for learning a downstream task. The generated query plan graph vector will condense important information about the physical query plan. How to accurately represent both structural and node information in the tree model’s output can directly impact the overall performance of the query optimizer. In this study, we focus on the representation capability of query plan tree models in the same application scenario using the same structure of feature encoder and cost estimator, so that the representation quality of a query plan can entirely rely on the tree model, allowing us to compare the performance of these tree models directly. 

Due to the tree structure characteristics of the query plan, it is difficult for tree models to use traditional machine learning methods to learn and aggregate node information directly, for they usually cannot accept inputs in the form of trees. Consequently, the database community has started developing machine-learning models designed explicitly for tree structures~\cite{treelstm,treecnn}. While the advanced models have marked performance improvements, they still have two main limitations:

\begin{itemize}
    \item \textbf{Information Dilution.} As information traverses from the leaf nodes—where specific relation-based operations are stored—towards the root, it tends to get diluted. This dilution process may negatively impact the accuracy of predictive cost estimations. The depth of the tree exacerbates this issue, leading to possible loss or dilution of critical information as it is passed upward or downward through the tree. The challenge of information dilution becomes more noticeable for complex queries that result in deeply structured trees.
    
    \item \textbf{Preserving Structural Information.} The plan tree contains essential structural and logical information. Preserving the integrity of this information while aggregating data from the entire graph is crucial and challenging.
    
\end{itemize}

Therefore, the design of the tree model used for processing query plans should consider the above two limitations to further improve the model's representation ability according to the structural characteristics of the query plan tree.

\section{Related Work}
Query plan representation is mainly studied under one of the following categories: studies proposing novel tree models and their comparative analysis.

\textbf{Tree models.} 
  RNN-based models like long short-term memory (LSTM)~\cite{sun2019end} and gated recurrent units (GRUs)~\cite{gru} are commonly used in query plan representation. Some works apply the self-attention mechanism to tree models, such as Saturn~\cite{saturn}, which aggregates LSTM's hidden layers based on attention as the query plan representation, or QueryFormer~\cite{queryformer}, which uses a Transformer~\cite{selfattention} to encode the query plan. However, these methods have to convert tree structure plans into sequential nodes as input, inevitably leading to the loss of tree structure information. 
 Machine learning models designed explicitly for tree structures, for instance, tree-structured LSTM (Tree-LSTM)~\cite{treelstm}, tree convolutional neural network (Tree-CNN)~\cite{treecnn}, allow information transfer between child nodes and parent node so that the learned node features can also contain structural information. While these approaches improve upon non-tree models, they still do not demonstrate good results in solving the problem of information dilution and effective aggregation of node features. Recently, some works~\cite{loger,roq} attempt to apply GNNs in this context. However, there is no GNN model specifically designed for query plan tree graphs yet; relevant works use the GNNs only to capture the query's join relationship in order to assist the representation learning by other tree models rather than being directly used for learning query plan trees.

\textbf{Comparative study.}
A significant contribution to this field is a study by Yao Z. et al.~\cite{2023comparative}. The researchers conducted a detailed analysis comparing the performance of mainstream feature encoders and tree models under various scenarios. However, their study asserts that the tree model has no obvious impact on the overall performance of the optimizer, which is not confirmed by the findings of the current research. Their analysis also has a limitation: it lacks a comprehensive assessment of tree models in the plan selection task, since their evaluation only relies on pairwise index selection~\cite{aimeetsai}, which does not adequately reflect the complexities in practical query optimization scenarios. Furthermore, they do not isolate the impact of the tree model architectures on this downstream task's performance.

\section{Model Architecture}
In this section, we first introduce the feature encoding method we used and then describe the structure of our proposed tree model BiGG.

\subsection{Feature Encoding}
\label{Feature Encoding}
\begin{figure}
\centering
\vspace {-0.15in}
\includegraphics[width=0.5\textwidth]{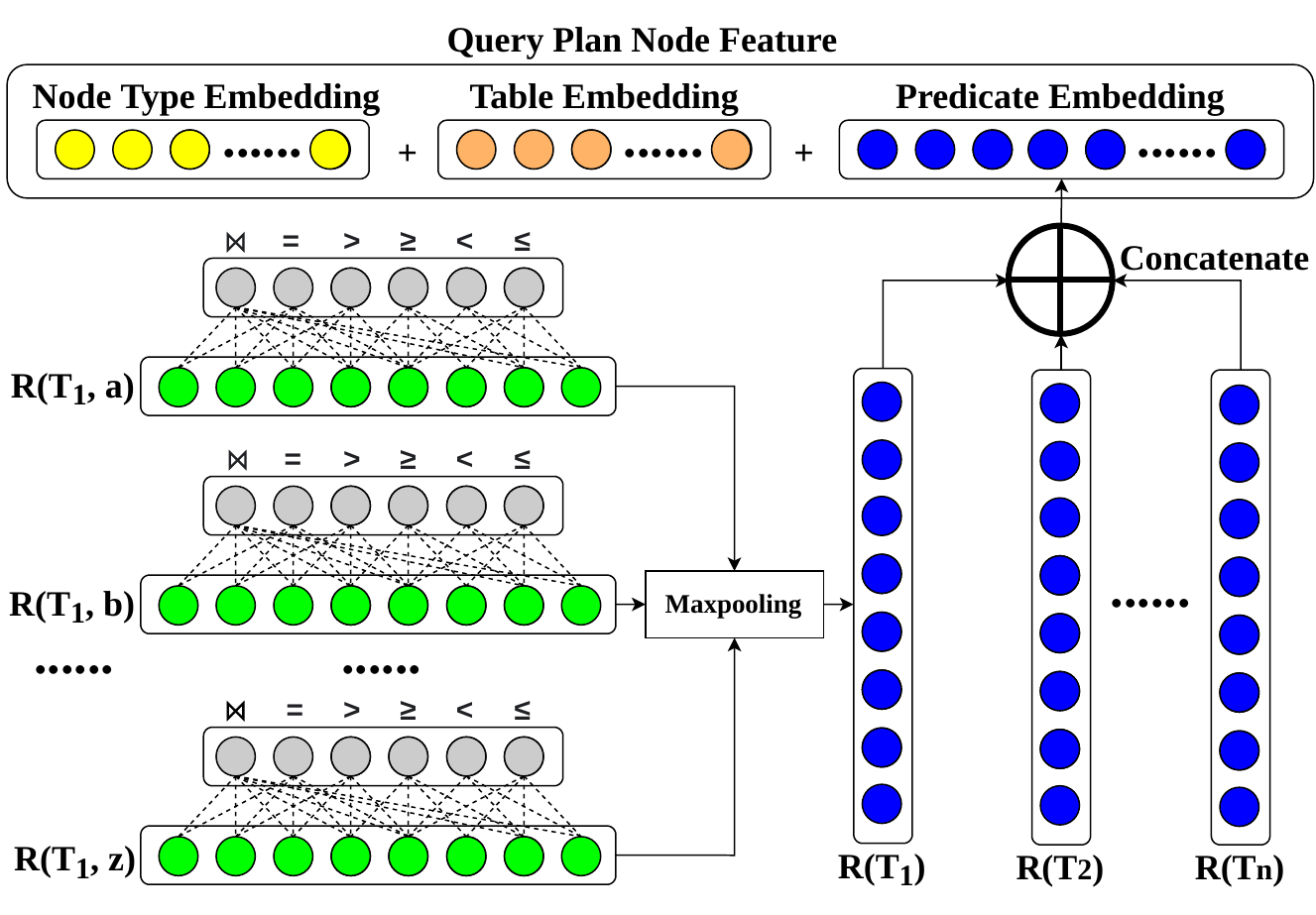}
\caption{Learning-based plan node representation.} \label{Query plan node feature}
\vspace {-0.15in}
\end{figure}

A query execution plan consists of information on the operators used to access and join data from various sources, including the physical implementation of the operators, their sequence, and the tables and columns accessed at each step. The amount of data flowing through each plan node, determined by the cardinality, is a decisive factor of the execution cost. In addition, the local and join predicates involved in the query plan may exhibit various levels of skewness and pairwise correlations, which in turn impact the accuracy of the cardinality estimates. To transform this complex information into fixed-length node features as input to the tree model, we developed a plan encoder inspired by RTOS \cite{rtos}, while making critical changes to suit our experimental purposes. As shown in Fig.~\ref{Query plan node feature}, each node feature in the query execution plan tree is composed of three parts: Node Type Embedding, Table Embedding, and Predicate Embedding.

\textbf{Node Type Embedding.} The node contains a specific operator type, such as Hash Join or Index Scan. We perform one-hot encoding on the node types and pass through a fully connected (FC) layer to obtain the node type embedding.

\textbf{Table Embedding.} We apply one-hot encoding on all tables in the node’s operation to obtain the Table Embedding.

\textbf{Predicate Embedding.} We consider the two main types of columns involved in the predicate: 

For numerical columns with values, the predicate operations in the node are classified into 6 cases: $\bowtie$, $=$, $>$, $\geq$, $<$, $\leq$. Each column is represented by a feature vector of the same length. If the predicate involves a specific operation on a certain column, the corresponding vector will be encoded according to the specific predicate type. If the column exists in a join predicate, the corresponding position will be 1 otherwise 0. For the other five cases, the predicate value is normalized to [0, 1] using the maximum and minimum values of the column in the database and plus 1 as the value of the corresponding position. If the predicate value exceeds the value range of the column in the database, according to the operator type, if there is no tuple that satisfies the predicate, the corresponding position is set to -1 otherwise 2. Uninvolved cases remain 0.

For other value-type columns, such as string type, it cannot be mapped to a value with interval meaning by a simple method (e.g. hash)~\cite{rtos}. Thus, the join predicate is encoded in the same way as above. For the other five cases, we use Word2vec~\cite{word2vec} to translate characters into numerical values for their respective positions in the column feature.

Each column has a dedicated FC layer to process its encoded feature vector and generate learned column embeddings. Then, max pooling is performed on all column embeddings belonging to the same table to obtain the embedding expression of each table with the same length. To avoid information loss during the aggregation of column embeddings in the node feature encoding stage, we directly concatenate embeddings of all tables as the predicate embedding.

\subsection{Bidirectional GNN for Query Plan Tree}
The basic idea of GNNs is that nodes can capture the characteristics and contextual information of neighbour nodes and their complex relationships via edges in the graph. Acknowledging the benefits of GNNs, due to query plans' characteristics discussed above, we propose a new tree model based on bidirectional GNN and GRU, which effectively addresses the following two main issues:

\begin{enumerate}
    \item Information Transmission between tree nodes
    \item Information Aggregation from node-level to graph-level  
\end{enumerate}    

\textbf{Information Transmission.} 
    In GNNs, message passing between nodes depends on both the existence of edges in graphs and their directionality. In the tree structure, this issue is mainly divided into two situations: single-directed and undirected edges. If single-directional edges are used, such as from the child node pointing to the parent node, information is always transmitted from the child node to the parent node. A node can only learn the information of its adjacent nodes in a GNN layer, which means any node in the tree will only learn information about its child nodes. When learning a deeper tree, for the information of the leaf nodes to be transferred to the root node, the same number of GNN layers as the depth of the tree is needed, which is generally unacceptable. The single-directed edge also makes it challenging to transfer the leaves' information to the root node, passing the entire graph without diluting or losing it, which makes using root node features as query plan graph-level representation infeasible. It also lets child nodes never learn the node feature and structural information about its parent node and above nodes, so the relevant structural information will inevitably be lost in the stage of aggregating node features. 

    Therefore, using undirected edges has the potential to significantly speed up the information transfer between nodes. However, our experiments have proven that simply using undirected edges cannot perform well in tree graphs. It destroys the structural relationship between parent and child nodes and makes the model unable to learn the hierarchical dependency between them. Inspired by DirGNNConv~\cite{dirgnn}, we innovatively treat the query plan trees as two one-way graphs with opposite edge directions. In each layer of our tree model BiGG, as shown in Fig.~\ref{Directed GNN}, we divide the original query plan into two tree graphs with opposite edge directions as input to two independent TransformerConv~\cite{transformerconv} layers respectively and integrate the corresponding nodes from the two output learned graphs through a learnable parameter, which makes it possible to transmit information in both directions while still utilizing the direction information of the edges and retaining relevant structural information. Given a query plan tree graph $G$ with $N$ nodes, $n$ as the nodes in the node feature set $Nodes$, $E$ as the undirected edge index set, $E_{\text{child to parent}}$ and $E_{\text{parent to child}}$ as the edge index sets with opposite directions, $p$ as a learnable parameter, the bidirectional GNN layer is defined as Algorithm~\ref{directed-gnn-algorithm}:
    \vspace {-0.15in}

\begin{figure}[t]
\centering
\includegraphics[width=\textwidth]{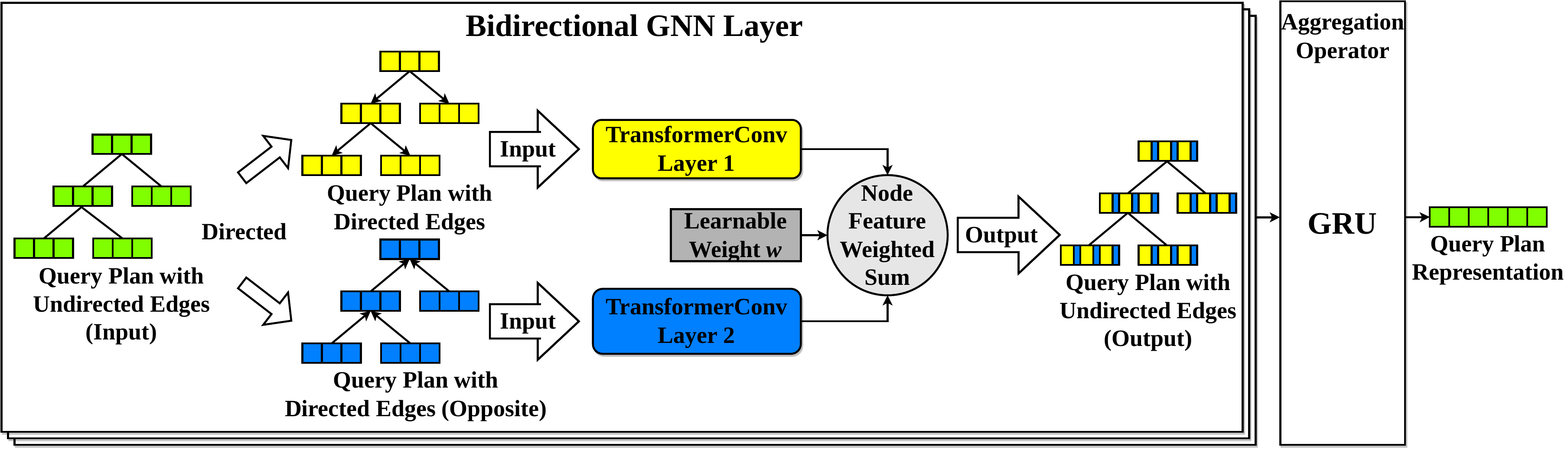}
\caption{Framework of tree model BiGG using bidirectional GNN aggregated by GRU.} \label{Directed GNN}
\vspace {-0.15in}
\end{figure}

\begin{algorithm}
\caption{Update node features in bidirectional GNN}
\label{directed-gnn-algorithm}
\begin{algorithmic}[1] 
\State \textbf{Input:} $G = (Nodes, E)$
\Statex $Nodes = \{n_i \mid \forall i \in \{1, \ldots, N\}\}$
\Statex $E = E_{\text{child to parent}} \cup E_{\text{parent to child}}$
\State $Nodes_{\text{child to parent}} = \text{TransformerConv1}(Nodes, E_{\text{child to parent}})$
\State $Nodes_{\text{parent to child}} = \text{TransformerConv2}(Nodes, E_{\text{parent to child}})$
\For{$i = 1$ to $N$}
    \State $n_{\text{new}_i} = p \times n_{\text{child to parent}_i} + (1-p) \times n_{\text{parent to child}_i}$
\EndFor
\State \textbf{Output:} $G = (Nodes_{\text{new}}, E)$
\end{algorithmic}
\end{algorithm}

\vspace {-0.15in}
    Additionally, the graph attention mechanism~\cite{selfattention} in TransformerConv can further improve the model's ability to capture the tree's local graph topology and global dependencies. To verify this perspective, our experiments also evaluated the influence of undirected edges, single-directed edges, and weighted directed edges on the model's capability to represent query plans.

\label{directed-gnn-aggregation}

\textbf{Information Aggregation.} 
    Conventional graph aggregation methods often produce inferior results when processing tree structure graphs. These methods usually simply aggregate node features by global pooling, which ignores the structural information in the graph. In our model we innovated by applying GRU to aggregate the GNN-learned query plan node features after post traversal of the tree. The rationale for this design decision is based on the observation that the order in which DBMS executes the query plan tree nodes and the order of nodes obtained by post-order traversal of the tree are similar~\cite{saturn}. This design learns the dependencies between nodes while more closely conforming to the actual execution sequence, thereby allowing our tree model to obtain the query plan's graph-level embedding while retaining the node features and structural information in the query plan tree to a certain extent. Our experiments have proven that GRU, as the aggregation operator, performs far better than other aggregation methods in query plan representation.

\section{Experimental Study}
Cost estimation is a crucial step in the query optimizer, as it directly determines how the model selects the optimal plan from candidate plans. However, these cost estimates are subject to arbitrary errors rooted in various sources, such as inaccurate cardinality estimates or simplifying assumptions. Therefore, in this work, we opt to learn the plan execution latency from runtime.

In addition to evaluating tree model cost estimation error and correlation, model performance in plan selection is also crucial. In the plan selection phase of query optimization, due to data and model uncertainties~\cite{roq}, the model's cost predictions may have inevitable errors, potentially affecting the final optimal plan selection. In practical applications, if a model can select the actual optimal plan is far more important than making more accurate predictions. Therefore, we evaluate the model's ability to correctly select the optimal plan with the shortest actual execution latency among multiple candidate query plans. We simulate real application scenarios to explore and compare the impact of employing different tree models on the overall plan selection accuracy of the optimizer.

To assess the impact of different tree models in query plan representation, we use the same feature encoding model described in Section~\ref{Feature Encoding} and the same multilayer perceptron (MLP)-based cost estimator for all experiments. In this manner, we can evaluate the representation ability of different tree models through the accuracy of cost estimates and plan selection.
\subsection{Experimental Setup}
The experiments are performed on a Linux server with 8-core Intel Silver 4216 Cascade Lake 2.1GHz CPU, 64GB memory, and a 32GB NVIDIA V100 Volta GPU. PostgreSQL 15.1 is used as the RDBMS for compiling and executing the workloads. The prototype code is written in Python 3.10 with the machine learning library PyTorch~\cite{pytorch}. All experimental results are the average after 10-fold cross-validation.

\textbf{Selected Existing Tree Models.} We select five state-of-the-art tree models for further evaluation as follows:

\begin{enumerate}
    \item 
        \textit{LSTM:} LSTM is a variant of recurrent neural network (RNN) architecture designed for sequences of data and capable of remembering long-term dependencies~\cite{sun2019end}. Since it cannot directly handle the tree, we flatten the query plan according to post-order traversal~\cite{saturn} as input. The hidden layer of the last node is treated as a graph-level representation of the query plan.
    \item 
        \textit{GRU:} GRU is a type of RNN that simplifies the LSTM architecture by combining the forget and input gates into an update gate, improving efficiency and performance on tasks involving sequential data~\cite{gru}. Model input and output are obtained in the same way as LSTM.
    \item 
        \textit{LSTM + Self-Attention:} Saturn~\cite{saturn} proposed an improvement of LSTM for query plan representation by applying the self-attention~\cite{selfattention} mechanism to weigh each hidden layer and aggregate all hidden layers according to their weight as the plan-level representation.
    \item 
        \textit{Tree-LSTM:} Tree-LSTM is an adaptation of the conventional RNN for tree-structured graphs, which can aggregate information from the leaf nodes to the root across different tree branches by generalizing the traditional LSTM cell by accepting inputs from multiple channels~\cite{treelstm}. The root node representation is treated as the graph-level representation of the query plan.
    \item 
        \textit{Tree-CNN:} Tree-CNN is designed to handle tree-structured data~\cite{treecnn}, and was first applied to query plan representation in NEO~\cite{neo}. It slides a triangular kernel from the root to the leaf nodes, which allows learning relationships between each combination of two child nodes and their parent node. Dynamic pooling is used to aggregate all features into a graph-level representation of the query plan~\cite{neo}. This model can only be applied to binary trees. Non-binary trees require preprocessing, involving adding more nodes and layers, to become compatible with the model.
\end{enumerate}

\vspace{-0.05in}

\label{dataset}
\textbf{Dataset.} We use workloads based on a 10 GB database generated by TPC-DS 3.2~\cite{tpcds}, an industrial standard benchmark commonly used to evaluate the performance of cost estimation with 25 tables and 429 columns. TPC-DS has more relations and allows for more complex query patterns compared with other benchmarks, such as IMDB (JOB)~\cite{imdb} and TPC-H~\cite{tpch}. Therefore, using TPC-DS we can generate more complex join queries to evaluate the representation learning performance of the tree models in more complex scenarios. Due to the limited complexity of the query generation template that comes with TPC-DS, we used a random query generator to generate queries based on TPC-DS relations. The queries are determined by parameters provided to the query generator, including the number of joins, join types (such as inner-join, outer-join, anti-join), the number of join and local predicates specified within \textit{Where} clauses along with the types of operators used. The following are the datasets used in the two tasks:
\begin{enumerate}
    \item 
        \textit{Dataset for cost estimation.} We generated 22k queries and each query has up to 10 joins, sampled from referential integrity one-to-many and artificial many-to-many joins, with each join having up to 3 join predicates and each table with up to 5 local predicates. PostgreSQL's default query plans for each query and their execution time are seen as samples and labels. 18k of the query plans are used for training, 2k for validation and 2k for testing.
    \item 
        \textit{Dataset for plan selection.} We generate 10k queries using the same query generator for evaluating plan selection performance. Each query was compiled in PostgreSQL using the 13 most influential optimization hints~\cite{roq} inspired by Bao~\cite{bao}. These hint sets introduced specific constraints on the join and access operators utilized within the query plans. Each sample in the dataset is a collection of candidate query plans generated based on the 13 hints from the same query. The dataset contains 10k queries yielding about 130k query plan trees and is divided into 8k queries for training, 1k for validation, and 1k for testing.
\end{enumerate}

\vspace{-0.05in}
\textbf{Evaluation Metrics.} We employ four evaluation metrics for the experiment, as shown in the following. Metrics 1-3 are applied to the cost estimation task, while metric 4 is dedicated to the plan selection task.

\vspace{-0.05in}
\begin{enumerate}
    \vspace{-0.05in}
    \item \textit{Prediction Error:} We use Q-Error to evaluate the accuracy of estimated latency. Given the estimated latency $y_{el}$ and actual latency $y_{al}$, the Q-Error is defined in Equation~\ref{qerror}, ranging from 1 (perfect accuracy) to $\infty$.
        \begin{equation}
            Q\text{-}Error = \frac{\max(y_{el}, y_{al})}{\min(y_{el}, y_{al})}
            \vspace{-0.01in}
            \label{qerror}
        \end{equation}
        
    \item \textit{Correlation:} We use Spearman’s rank correlation to measure the relationship between model-estimated latency and actual latency, with values closer to 1 indicating a stronger correlation. Unlike the widely used Pearson’s coefficient, Spearman correlation is not such sensitive to outliers and data scales for using a monotonic function, making it more suitable to evaluate indicators such as latency that may have large orders-of-magnitude differences.

    \item \textit{Inference Overheads:} We measured the average time the tree model used to predict a latency estimate for each encoded query plan during testing, which can be used to compare the computational overhead of each tree model.

    \item \textit{Plan Suboptimality:} Given a set of candidate execution plans $p$ generated based on the same query, ranking these candidate plans according to the actual execution latency $AL(.)$ and identifying the plan with the shortest execution latency as the actual optimal plan, denoted as $p_o$. Concurrently, the optimal plan selected by the optimizer with the shortest predicted latency is donated as $p_m$. In this context, we define Plan Suboptimality as Equation~\ref{plan_suboptimality} which takes a value in the range of $[1, \infty)$ and the value closer to 1 reflects the model's better ability to identify the optimal plan among multiple options.

    \begin{equation}
    \label{plan_suboptimality}
    \text{Plan Suboptimality} = \frac{AL(p_m)}{AL(p_o)}
    \vspace{-0.01in}
    \end{equation}

\end{enumerate}

\textbf{Loss Function.} We used the Mean Squared Error algorithm based on the cost estimator model's output and labels after being preprocessed by natural log transformation and min-max scaling. To reduce the significant skewness in execution latency values, we first apply a natural log to all labels and make them more suitable targets for a machine learning model to learn~\cite{roq}. Then, min-max scaling is performed to normalize the processed label to the range [0, 1], aligning it with the same output range of the Sigmoid activation function used in the final layer of the cost estimator. Given all the labels in the dataset after natural logarithmic transformation $y_{\log_e}$, a cost estimator output $y_{out}$, the corresponding actual latency $y_{al}$, the loss function is defined as follows.
\vspace{-0.025in}

\begin{equation}
\text{Loss} = \sum_{i=1}^{n} \left( \frac{\log_{e}(y_{{al}_i}) - \min(y_{\log_e})}{\max(y_{\log_e}) - \min(y_{\log_e})} - y_{{out}_i} \right)^2
\end{equation}

\textbf{Model Training and Parameter Tuning}. In the model training phase, the hyper-parameters are individually tuned for each tree model. The learning rate and other hyper-parameters of the remaining models, such as the feature encoder or cost estimator, are jointly tuned. Parameter tuning is done using the Asynchronous Successful Halving Algorithm (ASHA)~\cite{asha}. Adaptive Moment Estimation (ADAM)~\cite{adam} is used as the optimizer for model training. Dropout and Early stopping are used to avoid over-fitting.

\subsection{Existing Tree Model Cost Estimation Performance}

We evaluate the impact of using different tree models under the same framework on the overall cost estimation performance of the query optimizer. We present the results of the current state-of-the-art tree models in the upper part of Table~\ref{tree-models-cost-estimation}.

Even in workloads with larger scale and more complex join relationships, five existing tree models, including GRU, LSTM, LSTM + Self-Attention, Tree-LSTM, and TCNN, do not have a significant difference in query plan representation capabilities as stated in the study by Yao Z. et al.~\cite{2023comparative}. However, in the special scenario of representing complex query plans, the differences in Q-Error and Spearman's Correlation performance between different models become more apparent. We outline interesting observations regarding the impacts of various characteristics of the models on plan tree representation learning as follows.

\begin{figure}[t]
\centering
    \subfigure[]{
        \includegraphics[width=0.26\textwidth]{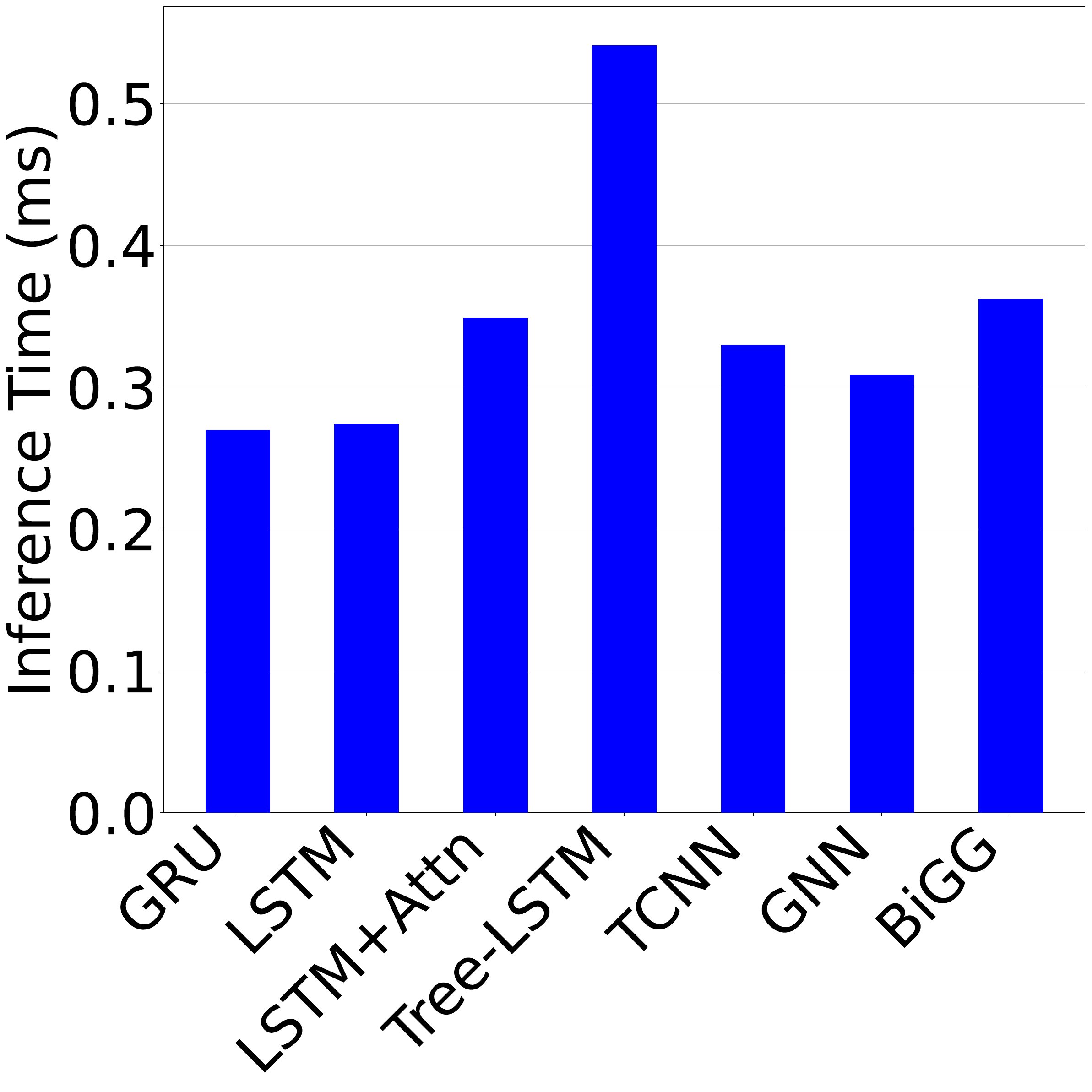}
        \label{Inference time}
        }
    \subfigure[]{
        \includegraphics[width=0.27\textwidth]{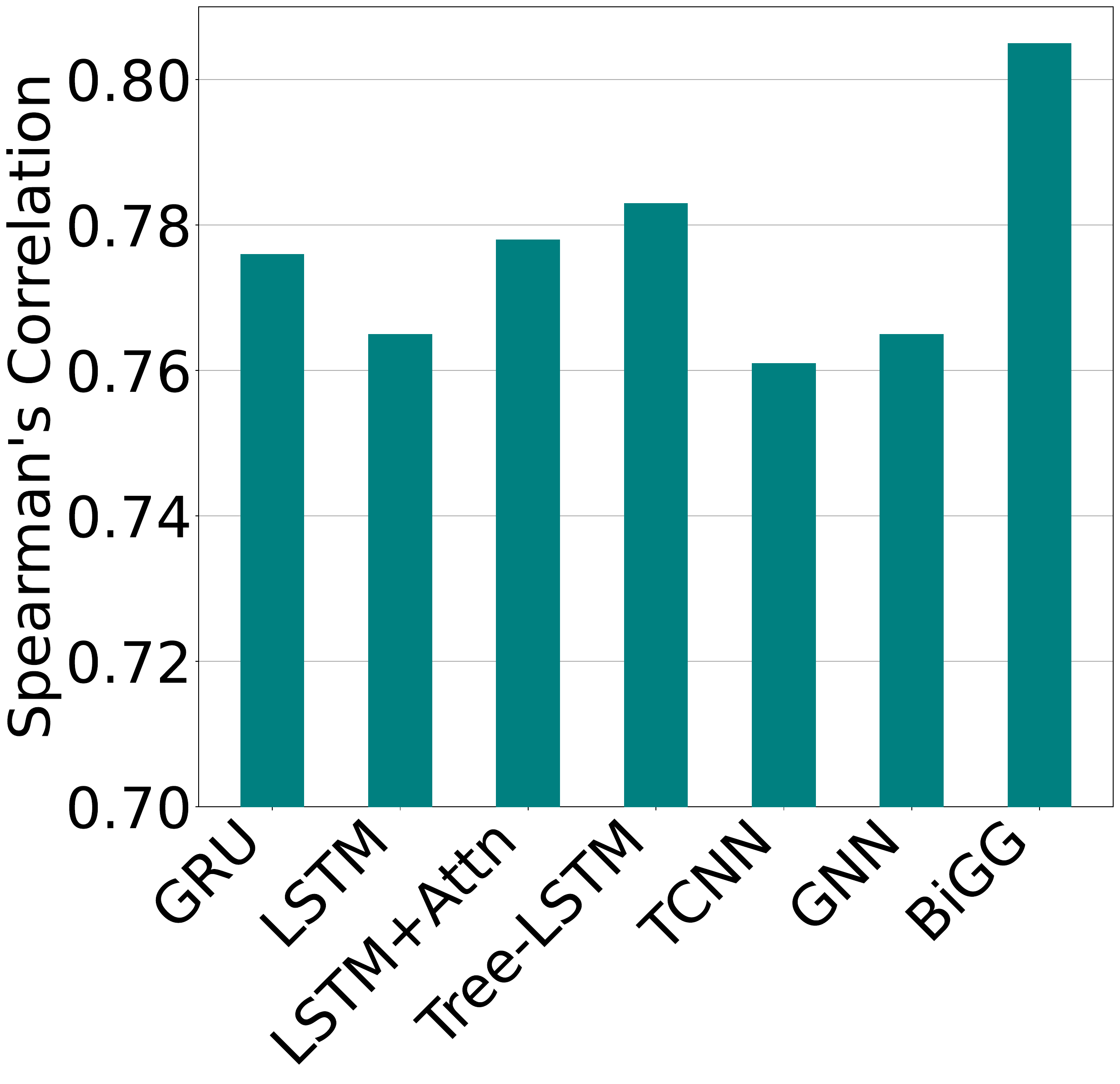}}
    \subfigure[]{
        \includegraphics[width=0.40\textwidth]{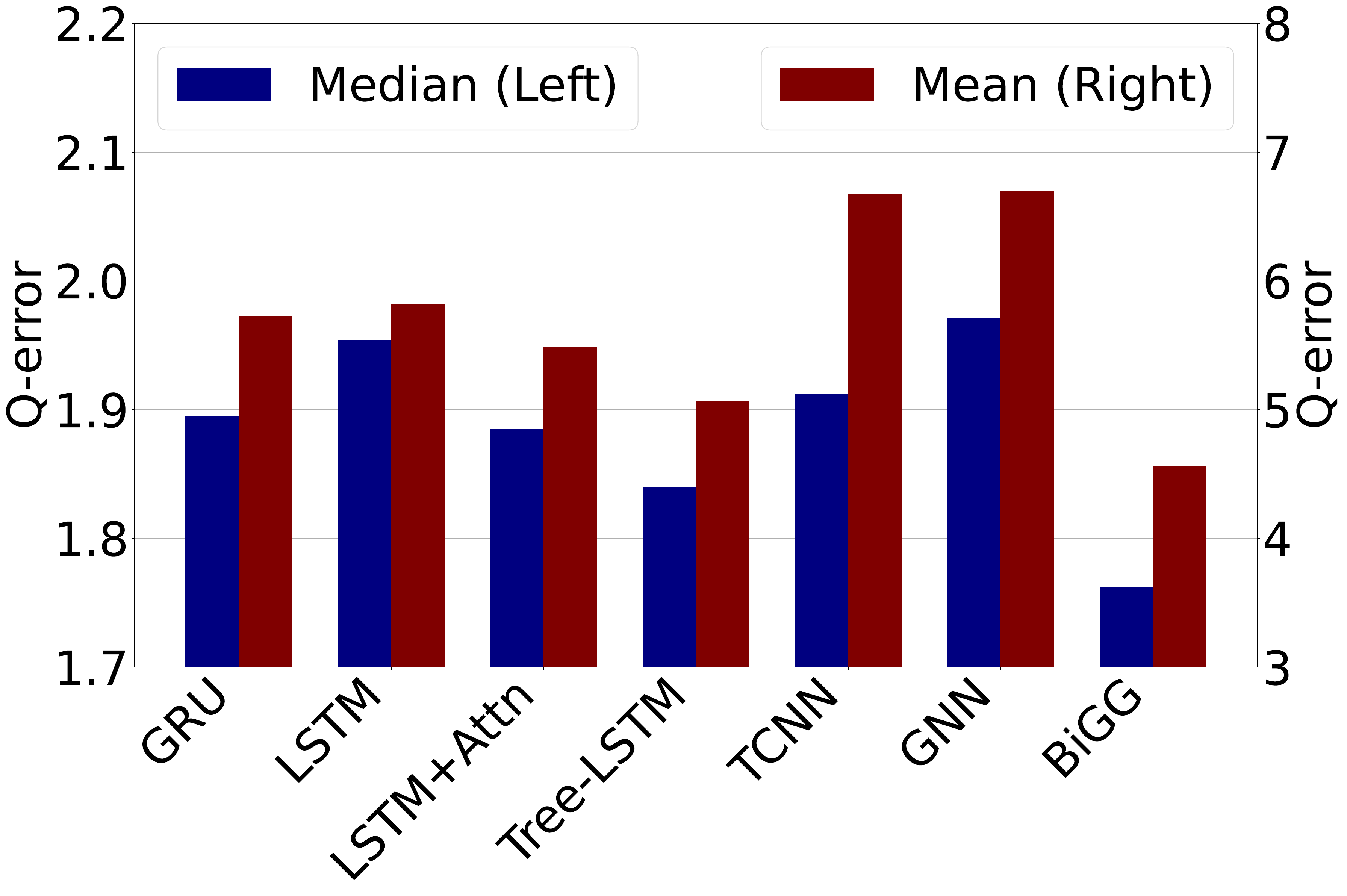}}
\vspace {-0.15in}
\caption{Performance comparison of different tree models of the cost estimation task in TPC-DS workload. (a) is the inference time of each tree model, (b) is Spearman's rank correlation coefficient and (c) is the mean and median Q-error experimental results.}
\label{cost estimation result}
\vspace {-0.2in}
\end{figure}

\textbf{Observation: Tree-LSTM has the best performance.} Compared with other models, Tree-LSTM performs best in all evaluation indicators. The highest Spearman's correlation and the lowest average Q-Error indicate that it can accurately capture and express critical information in most cases, even when processing complex query plans. In terms of Q-Error distribution, it has a significantly smaller tail-end error than other tree models, indicating that it is more robust in dealing with outliers and extreme situations. Experimental results demonstrate the effectiveness of its hierarchical structure-aware processing in accurately predicting complex query plan costs. However, this comes at the cost of having the highest inference overheads, as shown in Fig.~\ref{Inference time}.

\textbf{Observation: GRU has a better performance than LSTM.} GRU and LSTM have similar principles to capture long-term dependencies, and both demonstrate relatively good and stable performance in the experiment. However, we observe that GRU, which has a simpler architecture and fewer model parameters, has better performance than LSTM across all evaluation metrics. Due to the reduced overhead and enhanced performance, GRU may offer advantages in query-plan representation tasks. Due to its simplified design, it requires less training data than LSTM to learn the dependencies between query plan nodes. Additionally, its mechanism of reset gate and update gate~\cite{gru} may enable GRU to more effectively forget unimportant information while learning the query plan tree of post-order expansion, leading to more accurate predictions.

\textbf{Observation: Self-attention enhances LSTM.} Comparing the results of LSTM and LSTM+Self-Attention in the experiment, we observe that integrating LSTM and self-attention as a tree model has better cost estimation performance than the basic LSTM. This means that using the self-attention mechanism to aggregate hidden states of all nodes can more effectively aggregate node features and better represent the query plan than relying solely on the final node's hidden state. Such findings show the potential of self-attention mechanisms in improving the tree model, demonstrating their capability to enrich model understanding and performance in complex query plan representation tasks.

\begin{table}[t]
\centering
\caption{Cost Estimation Accuracy (Q-Error) in TPC-DS workload. The first block includes the state-of-the-art tree models we evaluated. The second block includes the tree models based on GNN layers with different graph edge directions and aggregation methods. The overall best results across blocks are underlined.}
\label{tree-models-cost-estimation}
\resizebox{\textwidth}{!}{
\begin{tabular}{|c|c|c|c|c|c|c|c|}
\hline
\makecell{\textbf{Tree Model (+Aggregation Method)}} & \makecell{\textbf{Graph Edge}\\\textbf{Direction}} & \makecell{\textbf{Median}\\\textbf{Q-error}} & \makecell{\textbf{90th}\\\textbf{Q-error}} & \makecell{\textbf{99th}\\\textbf{Q-error}} & \makecell{\textbf{Spearman’s}\\\textbf{Correlation}} & \makecell{\textbf{Top 50\%}\\\textbf{Mean Q-error}} & \makecell{\textbf{Top 99\%}\\\textbf{Mean Q-error}} \\
\hline
\textbf{GRU} & - & 1.895 & 22.582 & 296.242 & 0.776 & 1.378 & 5.727 \\
\textbf{LSTM} & - & 1.954 & 21.979 & 311.205 & 0.765 & 1.398 & 5.824 \\
\textbf{LSTM + Self-Attention} & - & 1.885 & 20.710 & 313.775 & 0.778 & 1.370 & 5.489 \\
\textbf{Tree-LSTM} & - & 1.840 & 19.413 & 235.727 & 0.783 & 1.352 & 5.063 \\
\textbf{TCNN} & - & 1.912 & 25.758 & 430.101 & 0.761 & 1.380 & 6.674 \\ \hline\hline
\textbf{GNN + AddPool} & Single directed & 1.971 & 25.184 & 471.242 & 0.765 & 1.403 & 6.697 \\
\textbf{GNN + GRU} & Single directed & 1.884 & 20.960 & 324.524 & 0.776 & 1.369 & 5.684 \\
\textbf{GNN + GRU} & Undirected & 1.880 & 20.320 & 298.524 & 0.775 & 1.371 & 5.409 \\
\textbf{Bidirectional GNN + AddPool} & Weighted directed & 1.882 & 23.387 & 459.604 & 0.771 & 1.370 & 6.502 \\
\textbf{Bidirectional GNN + GRU} & Weighted directed & \underline{\textbf{1.762}} & \underline{\textbf{16.537}} & \underline{\textbf{199.137}} & \underline{\textbf{0.805}} & \underline{\textbf{1.327}} & \underline{\textbf{4.558}} \\
\hline
\end{tabular}}
\vspace {-0.2in}
\end{table}

\vspace {-0.05in}
\subsection{GNN-based Tree Model Cost Estimation Performance}

We explore and evaluate the impact of using different edge structures and aggregation methods on the representation ability of GNN-based tree models. We present the experiment results of these models in the lower part of Table~\ref{tree-models-cost-estimation}.

\textbf{Observation: Using GRU as an aggregation method works better than conventional methods.} Compared with using global add pooling, the experiment results indicate a significant improvement in the performance of GNN models across all evaluation metrics by employing GRU as the aggregation operator. The improvement verifies our perspective discussed in Section~\ref{directed-gnn-aggregation}. Expanding the tree structure graph through post-order traversal enables GRU to learn dependencies between nodes in the actual order of how nodes are executed in the DBMS and aggregate the node features utilizing GRU’s gating mechanism. By leveraging the GRU for aggregation, the generated query plan representation can preserve more structural information and more valuable node features, thereby enhancing the model's ability to predict cost estimation accurately.

\textbf{Observation: Utilizing the direction of the edges in the query plan tree improves the representation performance of GNN models.} Our experiments demonstrate that simply using single-directed or undirected edges does not have a noticeable impact on the representation performance of the GNN-based tree models. However, models employing bidirectional GNN with weighted directed edges all show significant cost estimation performance improvements, which both promote message passing between nodes in the tree structure graph and allow nodes to learn the parent-child relationship through weighted directed edges, thereby enabling the model to capture more accurately structural and logical information of query plan even in complex workload representation tasks.

\textbf{Observation: Bidirectional GNN + GRU Aggregation Operator performs much better than other tree models.} Our experimental findings confirm that the novel architecture that we introduced, combining bidirectional TransformerConv and GRU as the aggregation operator, significantly outperforms other GNN-based as well as the state-of-the-art non-GNN tree models. Although it requires a higher inference overhead for running two complete GNN layers, it undoubtedly demonstrates the great potential and research value of GNN technology in query plan representation.

\subsection{Plan Selection Performance and Analysis}
We evaluate the impact of different tree models on plan selection performance in scenarios where PostgreSQL generates multiple candidate plans using different hints. The experimental results on Plan Suboptimality are shown in Table~\ref{tree-models-plan-subop}.

\begin{table}[ht]
\vspace{-0.15in}
\centering
\caption{Plan Suboptimality performance of tree models in TPC-DS workloads. The best results in each column are underlined.}
\label{tree-models-plan-subop}
\resizebox{\textwidth}{!}{
\begin{tabular}{|c|c|c|c|c|c|c|}
\hline
\makecell{\textbf{Tree Model}} & \makecell{\textbf{Median}\\\textbf{Plan Subopt}} & \makecell{\textbf{90th}\\\textbf{Plan Subopt}} & \makecell{\textbf{99th}\\\textbf{Plan Subopt}} & \makecell{\textbf{Top 50\%}\\\textbf{Mean Plan Subopt}} & \makecell{\textbf{Top 90\%}\\\textbf{Mean Plan Subopt}} & \makecell{\textbf{Top 99\%}\\\textbf{Mean Plan Subopt}} \\
\hline
\textbf{GRU} & 1.049 & 2.478 & 97.991 & 1.013 & 1.139 & 1.581 \\
\textbf{LSTM} & 1.052 & 2.447 & 87.656 & 1.013 & 1.143 & 1.637 \\
\textbf{LSTM + Self-Attention} & 1.051 & 2.593 & 102.315 & 1.013 & 1.146 & 1.881 \\
\textbf{Tree-LSTM} & 1.047 & 2.551 & 72.518 & 1.012 & 1.137 & 1.824 \\
\textbf{TCNN} & 1.054 & 2.610 & 93.157 & 1.014 & 1.161 & 1.923 \\
\textbf{GNN + AddPool} & 1.052 & 2.543 & 100.953 & 1.013 & 1.148 & 1.700 \\
\textbf{Bidirectional GNN + GRU} & \underline{\textbf{1.045}} & \underline{\textbf{2.094}} & \underline{\textbf{85.441}} & \underline{\textbf{1.011}} & \underline{\textbf{1.122}} & \underline{\textbf{1.516}} \\
\hline
\end{tabular}}
\vspace {-0.15in}
\end{table}

\textbf{Observation: The impact of tree models on the plan selection of the optimizer is not significant.} Experimental data shows that the impact of different tree models on the plan selection task of the optimizer is not obvious. Although the models employ various strategies to learn query plan representations, the choice between pure tree models does not result in a sharp contrast in the optimizer's ability to select the plan closest to the optimal plan. This subtle impact suggests that factors outside the tree model, such as the design of other components within the optimizer or the specific optimizations based on the workload’s characteristics, may also play a crucial role in the overall robustness and effectiveness of the optimizer.

\textbf{Observation: Improvements in cost estimation performance do not imply improvements in plan selection.} In the experimental results shown in Table 1, the tree model we proposed based on bidirectional GNN has a cost estimation performance significantly ahead of other tree models. Although this trend is still reflected in the plan selection evaluation metric, it can no longer widen the gap with the performance of other models. This indicates that merely enhancing the model's cost estimation accuracy may not obviously improve the optimizer's real-world performance.

\section{Conclusions and Future Work}
We conduct a comparison and analysis of the performance and mechanisms of the current tree models used in query plan representation in complex workloads under tasks of cost estimation and plan selection. Subsequently, we introduce an innovative tree model BiGG leveraging bidirectional GNN aggregated by GRU to learn query plan representations. Through extensive experiments, our proposed model has shown significant enhancements in accuracy for both cost estimation and plan selection compared to existing models. The next phase of our research will explore applying this model to a complete learning-based query optimizer and further advancing its performance and robustness in plan selection. 

\bibliographystyle{splncs04} 
\bibliography{references}

\end{document}